\documentstyle[aps,preprint,eqsecnum,tighten,floats,rotate,epsf,prc]{revtex}

\newbox\rotbox

\begin{document}

\preprint{\vbox{\it Submitted to Phys. Rev. C \hfill\rm TRI-PP-95-40}}

\title{\huge Valid QCD Sum Rules\\ for\\ Vector Mesons in Nuclear
Matter}

\author{\sc Xuemin Jin\footnote{E-mail:~jin@alph02.triumf.ca
{}~$\bullet$~ Telephone: (604) 222-1047 ext.\ 6446 ~$\bullet$~ Fax:
(604) 222-1074 } and
Derek~B.~Leinweber\footnote{
Present address: Dept.\ of Physics, Box 351560,
University of Washington, Seattle, WA 98195. \hfill\break
E-mail:~derek@phys.washington.edu ~$\bullet$~ Telephone:
(206)~616--1447 ~$\bullet$~ Fax: (206)~685--0635
WWW:~http://www.phys.washington.edu/$\sim$derek/Welcome.html}}

\address{TRIUMF, 4004 Wesbrook Mall, Vancouver, B. C., V6T 2A3,
Canada}

\maketitle

\begin{abstract}
QCD sum rules for vector mesons ($\rho$, $\omega$, $\phi$) in nuclear
matter are reexamined with an emphasis on the reliability of various
sum rules.  Monitoring the continuum contribution and the convergence
of the operator product expansion plays crucial role in determining
the validity of a sum rule. The uncertainties arising from less than
precise knowledge of the condensate values and other input parameters
are analyzed via a Monte-Carlo error analysis.  Our analysis leaves no
doubt that vector-meson masses decrease with increasing density. This
resolves the current debate over the behavior of the vector-meson
masses and the sum rules to be used in extracting vector meson
properties in nuclear matter.  We find a ratio of $\rho$-meson masses
of $m_{\rho^*}/m_\rho=0.78\pm 0.08$ at nuclear matter saturation
density.
\end{abstract}
%

\newpage
%
\section{Introduction}
\label{intro}

Whether the properties of vector mesons might change significantly
with increasing nuclear matter density is of considerable current
theoretical interest. This interest is motivated by its relevance to
the physics of hot and dense matter and the phase transition of matter
{}from a hadronic phase to a quark-gluon plasma at high density and/or
temperature. In particular, the modifications of vector-meson masses
in nuclear matter have been studied extensively.

At least three experimentally based studies have been cited as
supporting the picture of decreasing vector-meson masses in nuclear
matter.  These include the quenching of the longitudinal response
(relative to the transverse response) in quasi-elastic electron
scattering \cite{altemus80}, $(e,e'p)$ reactions\cite{reffay88}, and
the discrepancy between the total cross section in $K^+$-nucleus
scattering on $^{12}$C and that predicted from an impulse
approximation calculation using $K^+$-nucleon scattering amplitudes
(extracted from $K^+$-D elastic scattering)\cite{noble81,%
celenza85,brown89,soyeur93,brown88}.  More direct investigations of
vector-meson masses in the nuclear medium have also been proposed. One
proposal is to study dileptons as a probe of vector mesons in the
dense and hot matter formed during heavy-ion collisions
\cite{dil}. The dilepton mass spectra should allow one to reconstruct
the masses of vector mesons decaying electromagneticly.

Theoretical investigations of vector-meson masses in nuclear matter
have used various approaches and models that include the scaling
ansatz of Brown and Rho\cite{brown91}, Nambu--Jona-Lasinio
model\cite{bernard88}, Walecka
model\cite{asakawa92,hermann92,kurasawa88,tanaka91,%
caillon93,jean94,shiomi94}, quark model\cite{saito95}, and the QCD
sum-rule approach\cite{hatsuda92,asakawa93,asakawa94,koike95}.
Previous studies of vector mesons at finite density via the QCD
sum-rule approach have been made by Hatsuda and Lee\cite{hatsuda92},
and subsequently by Asakawa and Ko\cite{asakawa93,asakawa94}. It was
found that the vector-meson masses decrease with increasing
density. This finding is consistent with the scaling ansatz proposed
in Ref.~\cite{brown91}, the quark model\cite{saito95} and predictions
obtained from the Walecka model provided the polarizations of the
Dirac sea are included\cite{%
kurasawa88,tanaka91,caillon93,jean94,shiomi94}.

However, more recently, another QCD sum-rule analysis has shed
considerable doubt on the former conclusions. Ref.~\cite{koike95}
claims that the previous QCD sum-rule analyses of the in-medium
vector-meson masses are incorrect and that the vector-meson masses
should in fact increase in the medium. In response, Hatsuda, Lee and
Shiomi\cite{hatsuda95} have argued that the scattering-length approach
used in Ref.~\cite{koike95} is conceptually erroneous.  In this paper,
we reexamine the QCD sum rules for vector mesons in nuclear
matter. Our focus is on the reliability and validity of various sum
rules. We will show that careful consideration of the validity of the
sum rules used to extract the phenomenological results is crucial to
resolving this debate.

Taking the three momentum to be zero in the rest frame of the nuclear
medium, one can only obtain one (direct) sum rule in each vector
channel. By taking the derivative of this sum rule with respect to the
inverse Borel mass, one may get an infinite series of derivative sum
rules. Hatsuda and Lee\cite{hatsuda92} used the ratio of the first
derivative sum rule and the direct sum rule in their analysis while
Koike\cite{koike95} argued that one should use the ratio of the second
and first derivative sum rules. We point out that in practical
sum-rule applications, the derivative sum rules are much less reliable
than the direct sum rule and eventually become useless as the number
of derivatives taken increases. The ratio method used by both previous
authors does not reveal the validity of each individual sum rule, and
hence can lead to erroneous results.

QCD sum rules relate the phenomenological spectral parameters (masses,
residues, etc.) to the fundamental properties of QCD.  To maintain
the predictive power of the sum-rule approach, the phenomenological
side of the sum rule is typically described by the vector meson pole
of interest plus a model accounting for the contributions of all
excited states. By working in a region where the pole dominates the
phenomenological side, one can minimize sensitivity to the model and
have assurance that it is the spectral parameters of the ground state
of interest that are being determined by matching the sum rules. In
practice, these considerations effectively set an upper limit in the
Borel parameter space, beyond which the model for excited states
dominates the phenomenological side.

At the same time, the truncated OPE must be sufficiently convergent%
\footnote{Here and in the following, ``convergence'' of the OPE simply
means that the highest dimension terms considered in the OPE, with
their Wilson coefficients calculated to leading order in perturbation
theory, are small relative to the leading terms of the OPE.}
as to accurately describe the true OPE. Since the OPE is an expansion
in the inverse squared Borel mass, this consideration sets a lower
limit in Borel parameter space, beyond which higher order terms not
present in the truncated OPE are significant and important. Monitoring
OPE convergence is absolutely crucial to recovering nonperturbative
phenomena in the sum-rule approach, as it is the lower end of the
Borel region where the nonperturbative information of the OPE is most
significant. This information must also be accurate and this point
will be further illustrated in Sec.~\ref{analysis}.

In short, one should not expect to extract information on the ground
state spectral properties unless the ground state dominates the
contributions on the phenomenological side and the OPE is sufficiently
convergent. In this paper, we will analyze each individual sum rule
with regard to the above criteria. A sum rule with an upper limit in
Borel space lower than the lower limit is considered invalid.  As a
measure of the relative reliability of various sum rules we consider
the size of the regime in Borel space where both sides of the sum
rules are valid. In addition, the size of continuum contributions
throughout the Borel region can also serve as a measure of
reliability, with small continuum model contributions being more
reliable.

The uncertainties in the OPE are not uniform throughout the Borel
regime. These uncertainties arise from an imprecise knowledge of
condensate values and other parameters appearing in the OPE.  As such,
uncertainties in the OPE are larger at the lower end of the Borel
region. To estimate these uncertainties we adopt the Monte-Carlo error
analysis approach recently developed in Ref.~\cite{leinweber95}. In
turn, these uncertainties provide error estimates for the extracted
phenomenological spectral properties. This is the first systematic
study of uncertainties for in-medium hadronic properties.

In the following we will show in detail how the direct sum rule is
valid and the most reliable. The first derivative sum rule suffers
{}from a small Borel region of validity and relatively large continuum
model contributions throughout.  It is marginal at best and any
predictions from this sum rule are unreliable. Higher derivative sum
rules are found to be invalid.  Both the direct sum rule and the first
derivative sum rule lead to the same conclusion that vector-meson
masses decrease as the nuclear matter density increases.

This paper is organized as follows: In Sec.~\ref{sumrule}, we sketch
the finite-density sum rules for vector mesons and discuss the
reliabilities of various sum rules.  In Sec.~\ref{analysis}, the sum
rules are analyzed and the sum-rule predictions are presented and
discussed. Sec.~\ref{conclusion} is devoted to a conclusion.

\section{Finite-density sum rules}
\label{sumrule}

In this section, we briefly review the QCD sum rules for vector mesons
in nuclear matter. We focus on some issues raised in these sum rules
and refer the reader to the
literature\cite{hatsuda92,asakawa93,asakawa94} for more details of the
sum rules.

QCD sum rules for vector mesons at finite-density study the
correlator defined by
\begin{equation}
\Pi_{\mu\nu}(q)\equiv i \int d^4 x\, e^{iq\cdot x}
\langle\Psi_0|TJ_\mu(x)J_\nu(0)|\Psi_0\rangle\ ,
\end{equation}
where $|\Psi_0\rangle$ is the ground state of nuclear matter,
$T$ is the covariant time-ordering operator \cite{ITZ80},
and $J_\mu$ represents any of the three conserved vector currents
of QCD:
\begin{equation}
J_\mu^\rho\equiv\mbox{$1\over 2$}
(\overline{u}\gamma_\mu u-\overline{d}\gamma_\mu d)\ ,
\qquad
J_\mu^\omega\equiv\mbox{$1\over 2$}
(\overline{u}\gamma_\mu u+\overline{d}\gamma_\mu d)\ ,
\qquad
J_\mu^\phi\equiv\overline{s}\gamma_\mu s \ .
\label{current}
\end{equation}

The nuclear medium is characterized by the rest-frame nucleon density
$\rho_N$ and the four-velocity $u^\mu$.  We assume that the medium is
invariant under parity and time reversal.  Lorentz covariance and the
conservation of the currents imply that the correlator
$\Pi_{\mu\nu}(q)$ can be decomposed into two independent structures
multiplying two invariants, corresponding to the transverse and
longitudinal polarizations.  The medium modifications of these two
invariants are in general different. To keep our discussion succinct,
we follow the earlier works and take ${\bf q}=0$ in the rest frame of
the medium, $u^\mu=\{1,{\bf 0}\}$.  Since there is no specific spatial
direction, the two invariants are related and only the longitudinal
part, $\Pi_L= \Pi^\mu_\mu/(-3q^2)|_{{\bf q}=0}$, is needed.

All three currents under consideration are neutral currents. This
implies that both time orderings in the correlator correspond to the
creation or annihilation of the vector meson.  Accordingly, the
spectral function is necessarily an even function of the energy
variable. One can write the invariant function as
$\Pi_L(q_0^2)=\Pi^\mu_\mu(q_0^2)/(-3q_0^2)$,%
\footnote{
In the baryon case, the correlation function, considered
in the rest frame as a function of $q_0$, has both even and odd parts.
The reason is that, at finite baryon density, a baryon in medium
propagates differently than an antibaryon, yielding a correlation
function that is asymmetric in the energy variable
(see Refs.~\cite{furnstahl92,cohen95}).}
which satisfies the following dispersion relation%
\footnote{Here we have omitted the infinitesimal as we are only
concerned with large and space-like $q_0^2$.}  \cite{hatsuda92}
\begin{equation}
\Pi_L(Q^2\equiv -q_0^2)=\frac{1}{\pi}\int_0^\infty ds\,
\frac{{\rm Im}\,\Pi_L(s)}
{s+Q^2}+{\mbox{subtractions}}\ .
\label{dis-0}
\end{equation}
To facilitate our discussion of derivative sum rules, it is useful to
derive the following dispersion relation for $\Pi_L^{(n)}(Q^2)\equiv
(Q^2)^n\Pi_L(Q^2)$
\begin{equation}
\Pi_L^{(n)}(Q^2)=\frac{1}{\pi}\int_0^\infty ds\,
\frac{{\rm Im}\,\Pi_L^{(n)}(s)}
{s+Q^2}+{\mbox{subtractions}}\ ,\hspace{1cm}(n\geq 1)\ ,
\label{dis-gen}
\end{equation}
with ${\rm Im}\Pi_L^{(n)}(s)=(-1)^n s^n{\rm Im}\Pi_L(s)$.

For large $Q^2$, one can evaluate the correlator by expanding the
product of currents according to the operator product expansion
(OPE). The result can in general be expressed as
\begin{equation}
\Pi_L(Q^2)=-c_0\ln(Q^2)+{c_1\over Q^2}+{c_2\over Q^4}+{c_3\over Q^6}
           +\cdots\ ,
\label{ope-gen}
\end{equation}
where we have omitted the polynomials in $Q^2$, which vanish under
Borel transform. The first term corresponds to the perturbative
contribution and the rest are nonperturbative power corrections. The
coefficients, $c_0$, $c_1$, $c_2$, and $c_3$, have been given in
literature. For $\rho$ and $\omega$ mesons, one
has\cite{hatsuda92,asakawa93}
\begin{eqnarray}
& &c_0={1\over 8\pi^2}\left(1+{\alpha_s\over \pi}\right),
\hspace*{1.5cm} c_1=0,
\nonumber
\\*[7.2pt]
& &c_2=m_q\,\langle\overline{q}q\rangle_{\rho_N}
+{1\over 24}\,\langle{\alpha_s\over \pi} G^2\rangle_{\rho_N}
+{1\over 4}\,A_2^{u+d}\,M_N\,\rho_N\ ,
\nonumber
\\*[7.2pt]
& &
c_3=-{112\over 81}\,\pi\,\alpha_s\,\langle\overline{q}q
\rangle_{\rho_N}^2
-{5\over 12}\,A_4^{u+d}\,M_N^3\,\rho_N\ ,
\label{rho-coef}
\end{eqnarray}
where $\langle\hat{O}\rangle_{\rho_N}\equiv \langle\Psi_0|\hat{O}
|\Psi_0\rangle$ denotes the in-medium condensate, $m_q\equiv
{1\over 2}(m_u+m_d)$, $\langle\overline{q}q\rangle_{\rho_N}=
\langle\overline{u}u\rangle_{\rho_N}=
\langle\overline{d}d\rangle_{\rho_N}$, and
$A^{q}_{n}$ is defined as
\begin{equation}
A^q_n\equiv 2\int^1_0 dx\, x^{n-1}\left[
q(x,\mu^2)+(-1)^n\,\overline{q}(x,\mu^2)\right]\ .
\end{equation}
Here $q(x,\mu^2)$ and $\overline{q}(x,\mu^2)$ are the scale dependent
distribution functions for a quark and antiquark (of flavor $q$) in a
nucleon. We follow the standard linear density approximation to the
in-medium condensates,
$\langle\hat{O}\rangle_{\rho_N}=\langle\hat{O}\rangle_0+
\langle\hat{O}\rangle_N\,\rho_N$, with $\langle\hat{O}\rangle_0$ the
vacuum condensate and $\langle\hat{O}\rangle_N$ the nucleon matrix
element. The corresponding result for the $\phi$ meson
is\cite{asakawa94}
\begin{eqnarray}
& &c_0={1\over 4\pi^2}\,\left(1+{\alpha_s\over \pi}\right),
\hspace*{1.5cm} c_1=0,
\nonumber
\\*[7.2pt]
& &c_2=2\,m_s\,\langle\overline{s}s\rangle_{\rho_N}
+{1\over 12}\,\langle{\alpha_s\over \pi} G^2\rangle_{\rho_N}
+A_2^{s}\,M_N\,\rho_N\ ,
\nonumber
\\*[7.2pt]
& &
c_3=-{224\over 81}\,\pi\,\alpha_s\,\langle\overline{s}s
\rangle_{\rho_N}^2
-{5\over 3}\,A_4^{s}\,M_N^3\,\rho_N\ .
\label{phi-coef}
\end{eqnarray}

We adopt the usual pole plus continuum ansatz for the spectral
density.  This ansatz was recently tested in the Lattice QCD
investigation of Ref.~\cite{leinweber95a} where it was found to
describe nucleon correlation functions very well.  The
phenomenological spectral density for vector mesons in medium takes
the form\cite{hatsuda92,asakawa93,asakawa94}
\begin{equation}
{1\over \pi}\,{\rm Im}\Pi_L(s)={\rho_{\rm sc}\over 8\pi^2}\,\delta(s)
+F^*_{\rm v}\,\delta\left(s-m_{\rm v}^{*2}\right)
+c_0\,\theta(s-s_0^*)\ ,
\label{ansatz}
\end{equation}
where the first term denotes the contribution of the Landau
damping, $m_{\rm v}^*$ is the vector-meson mass in the medium,
and $s^*_0$ is the continuum threshold. In the calculations to
follow, we use $m_{\rm v}$, $F_{\rm v}$ and $s_0$ to denote
the corresponding vacuum (zero density limit) parameters.

Substituting Eqs.~(\ref{ope-gen}) and (\ref{ansatz}) into the
dispersion relation of~({\ref{dis-0}) and applying the Borel
transform to both sides, one obtains the direct sum rule
\begin{eqnarray}
F^*_{\rm v}\, e^{-m_{\rm v}^{*2}/M^2}&=&-{\rho_{\rm sc}\over 8\pi^2}
+c_0\,M^2\,E_0
\nonumber
\\*[7.2pt]
&+&c_1+{c_2\over M^2}+{c_3\over 2!\, M^4}+\cdots
+{c_m\over (m-1)!\,(M^2)^{m-1}}+\cdots\ .
\label{sum-gen-0}
\end{eqnarray}
{}From the dispersion relation of~(\ref{dis-gen}), one finds
\begin{eqnarray}
F^*_{\rm v}\, (m_{\rm v}^{*2})^n \, e^{-m_{\rm v}^{*2}/M^2}&=&
n!\, c_0 \,(M^2)^{n+1}\,E_n
\nonumber
\\*[7.2pt]
&+&(-1)^n\left[c_{n+1}+{c_{n+2}\over M^2}
+\cdots+{c_{m}\over (m-n-1)!\,(M^2)^{m-n-1}}+\cdots\right] \ ,
\label{sum-gen-n}
\end{eqnarray}
where we have defined
\begin{equation}
E_k\equiv 1-e^{s_0^*/M^2}\left[\sum_{i=0}^k
{1\over (k-i)!}\left({s_0^*\over M^2}\right)^{k-i}\right]\ .
\end{equation}
One recognizes that Eq.~(\ref{sum-gen-n}) corresponds
to the derivative sum rules as they may also
be obtained by taking derivatives of Eq.~(\ref{sum-gen-0})
with respect to $1/M^2$.

Hatsuda and Lee used Eq.~(\ref{sum-gen-0}) and the first derivative
sum rule ($n=1$) of~(\ref{sum-gen-n}) in their
calculations\cite{hatsuda92}, while Koike claimed that one should use
the first and second ($n=2$) derivative sum rules\cite{koike95}.  Of
course, if one could carry out the OPE to arbitrary accuracy and use a
spectral density independent of the model for excited state
contributions, the predictions based on Eq.~(\ref{sum-gen-0}) and
those based on the derivative sum rules should be the same.  In
practical calculations, however, one has to truncate the OPE and use a
simple phenomenological ansatz for the spectral density. Thus it is
unrealistic to expect the sum rules to work equally well. The question
is, which sum rules give the most reliable predictions. To answer this
question, let us compare the $n^{\rm th}$ derivative sum rule with the
direct sum rule of~(\ref{sum-gen-0}). We observe the following:
\begin{enumerate}
\item The perturbative contribution in the derivative sum rule has an
extra factor $n!$ relative to the corresponding term in
Eq.~(\ref{sum-gen-0}), implying that the perturbative contribution is
more important in the derivative sum rules than in the direct sum
rule, and becomes increasingly important as $n$ increases.  Since the
perturbative term mainly contributes to the continuum of the spectral
density, maintaining dominance of the lowest resonance pole in the sum
rule will become increasingly difficult as $n$ increases.
\item In Eq.~(\ref{sum-gen-0}), the term proportional to $c_{m}$ is
suppressed by a factor of $1/(m-1)!$, while it is only suppressed by
$1/(m-n-1)!$ in the derivative sum rule ($m\,>\,n$). This implies that
the convergence of the OPE is much slower in the derivative sum rule
than in the direct sum rule. This arises because the convergence of
the OPE for $\Pi^{(n)}_L(Q^2)= (Q^2)^n\,\Pi_L(Q^2)$ is obviously much
slower than that for $\Pi_L(Q^2)$ for large $Q^2$.  Consequently, the
high order power corrections are more important in the derivative sum
rule than in Eq.~(\ref{sum-gen-0}), and become more and more important
as $n$ increases. If one would like to restrict the size of the last
term of the OPE to maintain some promise of OPE convergence, the size
of the Borel region in which the sum rules are believed to be valid is
restricted.
\item The power corrections proportional to $c_1, c_2, \cdots, c_n$ do
{\it not} contribute to the $n^{\rm th}$ derivative sum rule but do
contribute to Eq.~(\ref{sum-gen-0}) \cite{mentioned}. If one truncates
the OPE, part or all of the nonperturbative information will be lost
in the derivative sum rules. It is also worth noting that the leading
power corrections are the most desirable terms to have.  They do not
give rise to a term in the continuum model and they are not the last
term in the OPE, whose relative contribution should be restricted to
maintain OPE convergence.
\end{enumerate}
In practice, the predictions based on the direct sum rule
of~(\ref{sum-gen-0}) are more reliable than those from the derivative
sum rules, which become less and less reliable as $n$ increases. This
can also be demonstrated by analyzing the sum rules numerically.

\section{Sum-rule analysis and discussion}
\label{analysis}

If the sum rules were perfect, one would expect that the two sides of
the sum rules overlap for all values of the auxiliary Borel parameter
$M$. As mentioned above, one has to truncate the OPE and use a
phenomenological model for the spectral density in practical
calculations.  Hence, the two sides of the sum rules overlap only in a
limited range of $M$ (at best).

All the previous works have used the ratio method of two sum rules.
There, one chooses the continuum threshold to make the ratio of the
two sum rules as flat as possible as a function of Borel mass (the
residue $F^*_{\rm v}$ drops out in the ratio). Although it has also
been used in various sum-rule calculations in vacuum, we note that the
ratio method has certain drawbacks. First, the ratio method does not
check the validity of each individual sum rule.  It may happen that
individual sum rules are not valid while their ratio is flat as
function of Borel mass. Secondly, the ratio method cannot account for
the fact that sum rules do not work equally well.  The Borel region
where a sum rule is valid can vary from one sum rule to
another. Finally, the continuum contributions to the sum rules are not
monitored in the ratio method. If the continuum contribution is
dominant in a sum rule, one should not expect to get any reliable
information about the lowest resonance.

\subsection{Outline of the method}

To overcome these shortcomings of the ratio method, we adopt here the
optimizing procedure originated in Ref.~\cite{leinweber90}, which has
been extensively used in analyzing various vacuum sum
rules\cite{jin93} and finite-density sum rules\cite{furnstahl92}.  In
this method, one optimizes the fit of the two sides of each individual
sum rule in a fiducial Borel region, which is chosen such that the
highest-dimensional condensates contribute no more than $\sim 10\%$ to
the QCD side while the continuum contribution is less than $\sim 50\%$
of the total phenomenological side (i.e., the sum of the pole and the
continuum contribution). The former sets a criterion for the
convergence of the OPE while the latter controls the continuum
contribution. While the selection of $50\%$ is obvious for pole
dominance, the selection of $10\%$ is a reasonably conservative
criterion that has not failed in practice.%
\footnote{Reasonable alternatives to the $10\%$ and $50\%$ criteria
are automatically explored in the Monte-Carlo error analysis, as the
condensate values and the continuum threshold change in each sample.}
This point is further illustrated in the discussions to follow.  The
sum rule should be valid in this fiducial Borel region as the pole
contribution dominates the phenomenological side and the QCD side is
reliable. We then select $51$ points in the fiducial region and use a
$\chi^2$ fit to extract the spectral parameters.  The reader is
referred to Refs.~\cite{leinweber90,leinweber95} for more details of
the method.

Since QCD sum rules relate the spectral parameters to the properties
of QCD, any imprecise knowledge of the condensates and related
parameters will give rise to uncertainties in the extracted spectral
parameters. These uncertainties have not been analyzed systematically
in the previous works.  Here we follow Ref.~\cite{leinweber95} and
estimate these uncertainties via a Monte-Carlo error analysis.
Gaussian distributions for the condensate values and related
parameters are generated via Monte Carlo.  The distributions are
selected to reflect the spread of values assumed in previously
published QCD sum-rule analyses and uncertainties such as the
factorization hypothesis.  These distributions provide a distribution
for the OPE and thus uncertainty estimates for the QCD side which will
be used in the $\chi^2$ fit. In fitting the sum rules taken from the
samples of condensate parameters one learns how these uncertainties
are mapped into uncertainties in the extracted spectral parameters.

As in the previous works\cite{hatsuda92,asakawa93,asakawa94}, we
truncate the OPE at dimension six and keep only the terms considered
in the literature. In the linear density approximation, the quark and
gluon condensates can be written as
\begin{eqnarray}
&
&m_q\,\langle\overline{q}q\rangle_{\rho_N}=m_q\,
\langle\overline{q}q\rangle_0
+{\sigma_N\over 2}\,\rho_N\ ,
\\*[7.2pt]
& &m_s\,\langle\overline{s}s\rangle_{\rho_N}=m_s\,
\langle\overline{s}s\rangle_0
+y\,{m_s\over m_q}\,{\sigma_N\over 2}\rho_N\ ,
\\*[7.2pt]
& &\langle{\alpha_s\over\pi}G^2\rangle_{\rho_N}=
\langle{\alpha_s\over\pi}G^2\rangle_0
+\langle{\alpha_s\over\pi}G^2\rangle_N\,\rho_N\ ,
\end{eqnarray}
where $y\equiv \langle\overline{s}s\rangle_N/
\langle\overline{q}q\rangle_N$.  The values of vacuum condensates we
use are $a=-4\pi^2\, \langle\overline{q}q \rangle_0=0.62\pm
0.05\,\text{GeV}^3$\cite{leinweber95}, $b=4\pi^2\,
\langle(\alpha_s/\pi)G^2 \rangle_0=0.4\pm
0.15\,\text{GeV}^4$\cite{leinweber95}, and
$\langle\overline{s}s\rangle_0/\langle\overline{q}q\rangle_0 =0.8\pm
0.2$\cite{reinders85,leinweber90}.  The quark mass $m_q$ is chosen to
satisfy the Gell-Mann--Oakes--Renner relation,
$2\,m_q\,\langle\overline{q}q\rangle_0=-m_\pi^2\, f_\pi^2$, and the
strange quark mass is taken to be $m_s=(26\pm
2.5)\,m_q$\cite{gasser91}. We adopt $\sigma_N=45\pm
7\,\text{MeV}$\cite{gasser91},
$\langle(\alpha_s/\pi)G^2\rangle_N=-650\pm
150\,\text{MeV}$\cite{cohen95}, and $y=0.2\pm
0.1$\cite{hatsuda92}. For the moments of the parton distribution
functions, $A^q_n$'s, we quote the values given in
Ref.~\cite{hatsuda92} and assign a $20\%$ uncertainty to each value,
$A^{u+d}_2=0.9\pm 0.18$, $A^{u+d}_4=0.12\pm 0.024$, $A^s_2=0.05\pm
0.01$, $A^s_4=0.002\pm 0.0004$.  The strong coupling constant is taken
to be $\alpha_s/\pi=0.117\pm 0.014$ at $1\,\text{GeV}$
scale\cite{leinweber95}.

The values of both vacuum and in-medium four-quark condensates are not
well determined.  Early arguments placed the values of vacuum
four-quark condensates within $10\%$ of the vacuum factorized values
\cite{svz79}.  However, later analyses suggested that factorization
underestimates the four-quark condensates
significantly\cite{govaerts87,dominguez88}.  Parameterizing the
condensate as $\kappa\langle\overline{q}q\rangle_0^2$, we will
consider values of $\kappa=2\pm 1$ and $1.0\leq\kappa\leq
3.5$\cite{govaerts87,leinweber95}.  As for the in-medium four-quark
condensates, previous authors have adopted the in-medium factorized
values (mean field approximation)\cite{hatsuda92,%
asakawa93,asakawa94}. In the study of finite-density baryon sum
rules\cite{furnstahl92,cohen95}, it was found that the in-medium
factorized values of certain four-quark condensates led to results in
contradiction with experiment. However, it should be pointed out that
the four-quark operators appearing in the baryon sum rules are
different from those in the vector meson sum rules.  Here we
parameterize the in-medium four-quark condensates as
$\kappa\,\langle\overline{q}q\rangle_{\rho_N}^2$.

The Gaussian distributions for the condensate values and various
parameters are generated using the values given above.  The error bars
in the extracted fit parameters (see below) are given by the standard
deviation of the distribution after $100$ condensate values generated
via Monte Carlo.  It is perhaps worth emphasizing that the error bars
do not represent the standard error of the mean, which is $10$ times
smaller for this case.  Hence the error bars are representative of the
spread of input parameter values.  In addition, the uncertainty
estimates become insensitive to the number of Monte Carlo samples
after about $50$ samples. We normalize all finite-density spectral
parameters ($m^*_{\rm v}$, $F^*_{\rm v}$, and $s^*_0$) to their
corresponding values in vacuum (i.e., zero density limit).  Thus, the
error bars in the ratios are dominated by the uncertainties in the
density dependent terms of the in-medium condensates since the errors
in the vacuum sum rules and finite-density sum rules are correlated.

\subsection{Numerical results}

Let us start with the sum rules for $\rho$ and $\omega$ mesons. We
first analyze the direct sum rule of Eq.~(\ref{sum-gen-0}). The Landau
damping contribution proportional to $\rho_{sc}$ is very small at the
densities considered here\cite{hatsuda92}. Treating $\rho_{sc}$ as a
search parameter, we find that the direct sum rule predicts a value
for $\rho_{sc}$ in accord with the Fermi-gas approximation,
$\rho_{sc}\simeq 2\pi^2\rho_N/M_N$\cite{hatsuda95,hatsuda92}.  In
calculating the uncertainty of this parameter we found that there is
insufficient information in the sum rules to reliably determine this
small contribution. As a result we use the Fermi-gas relation and
treat $m^*_{\rm v}$, $F^*_{\rm v}$, and $s^*_0$ as search parameters
in the following.

\begin{figure}[b]
\begin{center}
\epsfysize=11.6truecm
\leavevmode
\setbox\rotbox=\vbox{\epsfbox{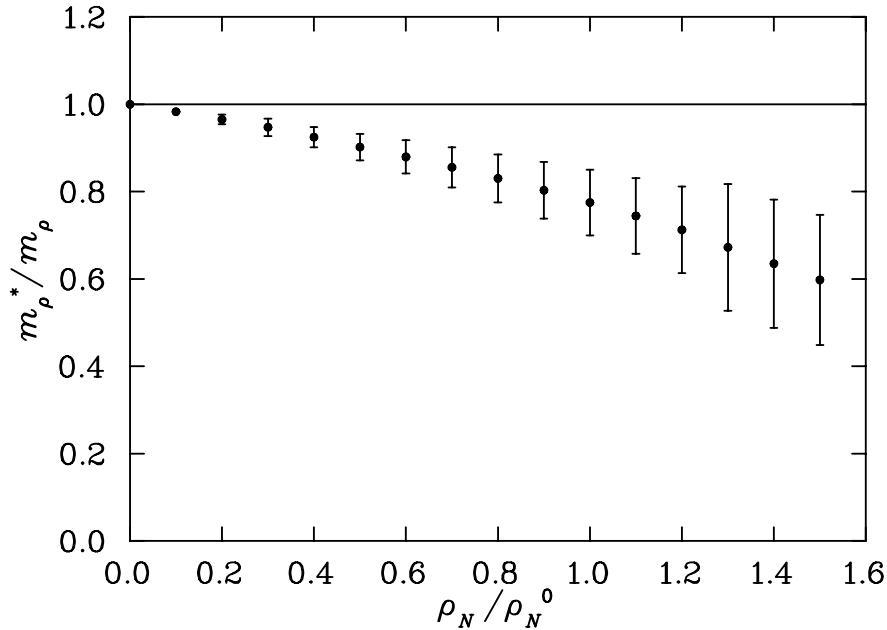}}\rotl\rotbox
\end{center}
\caption{Predictions of the direct sum rule for $m_\rho^*/m_\rho$ as a
function of the medium density.}
\label{fig-1}
\end{figure}

The predictions for the ratio $m_{\rho}^*/m_\rho$ as a function of the
nucleon density is plotted in Fig.~\ref{fig-1}. One can see that the
$\rho$-meson mass decreases with increasing density. At nuclear matter
saturation density $\rho_N=\rho^0_N=(110\,\text{MeV})^3$, we find
$m_{\rho}^*/m_\rho= 0.78\pm 0.08$.  The residue $F_\rho^*$ and the
continuum threshold $s_0^*$ also decrease as the density
increases. The predictions for the ratios $F^*_\rho/F_\rho$ and
$s_0^*/s_0$ are shown in Figs.~\ref{fig-2} and \ref{fig-3},
respectively.

\begin{figure}[p]
\begin{center}
\epsfysize=11.6truecm
\leavevmode
\setbox\rotbox=\vbox{\epsfbox{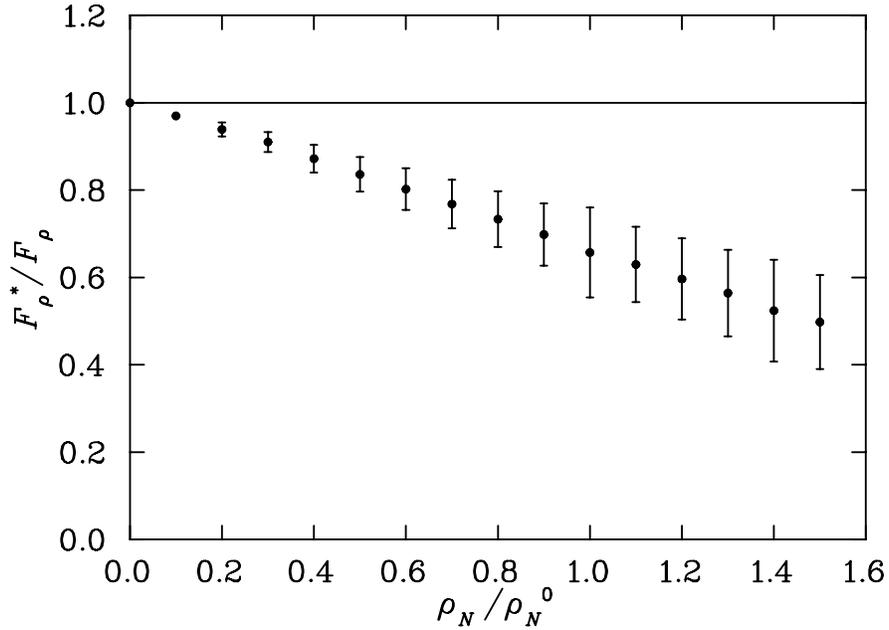}}\rotl\rotbox
\end{center}
\caption{Predictions of the direct sum rule for the ratio
$F_\rho^*/F_\rho$ as a function of the medium density.}
\label{fig-2}
\end{figure}

\begin{figure}[p]
\begin{center}
\epsfysize=11.6truecm
\leavevmode
\setbox\rotbox=\vbox{\epsfbox{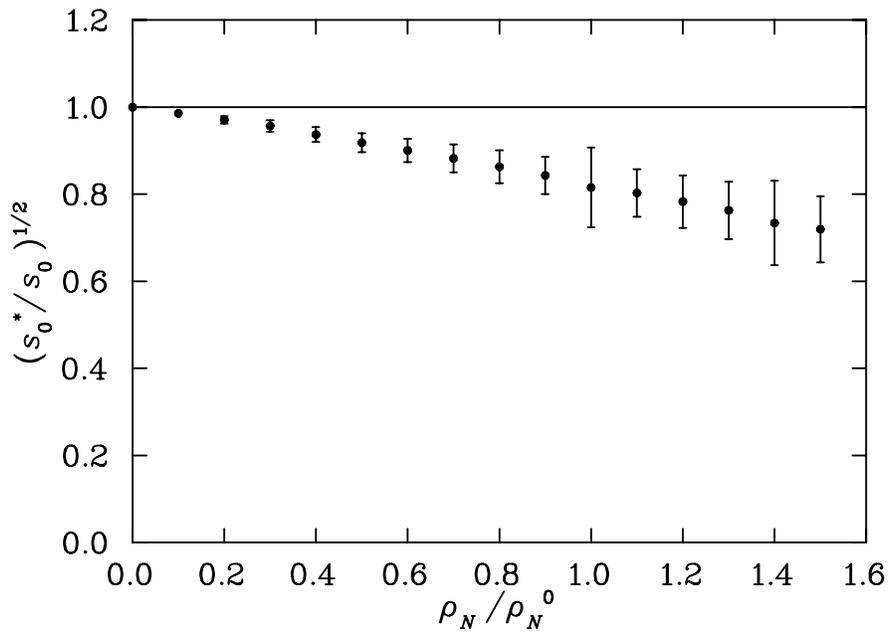}}\rotl\rotbox
\end{center}
\caption{Predictions of the direct sum rule for the ratio
$(s_0^*/s_0)^{1/2}$ as a function of the medium density.}
\label{fig-3}
\end{figure}

In Fig.~\ref{fig-4} the left- and right-hand sides of the direct sum
rule are plotted as functions of $M$ at the nuclear matter saturation
density.  The near perfect overlap of the two sides of this sum rule
is typical of the quality of fits seen at other densities.  The
corresponding valid Borel window and the relative contributions of the
continuum and the highest order term in the OPE are displayed in
Fig.~\ref{fig-5} as dashed curves. One notices that the direct sum
rule is valid in a broad Borel regime, where the highest order term
contributes less than $10\%$ and the continuum contributes only about
$15\%$ at the lower bound and the required $50\%$ at the upper
bound. Thus, the pole contribution truly dominates the sum rule in the
Borel region of interest, implying that the predictions are
reliable. We also find that both lower and upper bounds are functions
of the density and decrease as the density increases. The rate of
decrease for the upper bound is larger than that for the lower bound,
which means that the optimal Borel window shrinks with increasing
density.

\begin{figure}[t]
\begin{center}
\epsfysize=11.6truecm
\leavevmode
\setbox\rotbox=\vbox{\epsfbox{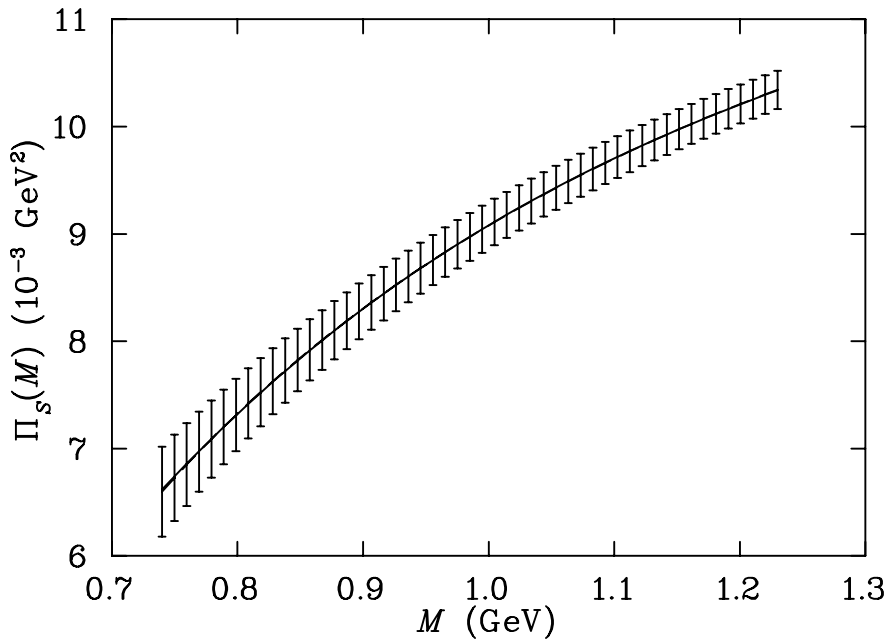}}\rotl\rotbox
\end{center}
\caption{The left- and right-hand sides of the direct sum rule as
functions of the Borel mass $M$ at the nuclear matter saturation
density.}
\label{fig-4}
\end{figure}

\begin{figure}[t]
\begin{center}
\epsfysize=11.6truecm
\leavevmode
\setbox\rotbox=\vbox{\epsfbox{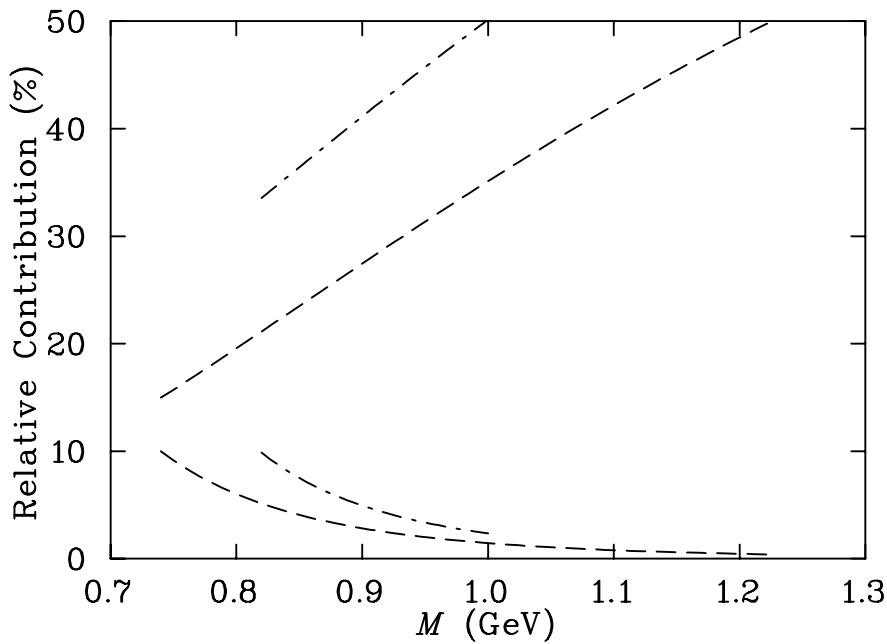}}\rotl\rotbox
\end{center}
\caption{Relative contributions of the continuum and the highest order
OPE terms to the sum rules as functions of the Borel mass at nuclear
matter saturation density. The dashed and dot-dashed curves correspond
to the direct sum rule and the first derivative sum rule,
respectively. Note the relatively broad regime of validity for the
direct sum rule.}
\label{fig-5}
\end{figure}

We proceed now to analyze the first derivative sum rule of
Eq.~(\ref{sum-gen-n}) (with $n=1$). It is found that this sum rule is
valid in a much smaller Borel regime.  The continuum and highest order
OPE term contributions are shown in Fig.~\ref{fig-5} as dot-dashed
curves for $\rho_N=\rho^0_N$.  It can be seen that the continuum
contribution exceeds $33\%$ in the entire Borel window and the
relative importance of the highest order term increases as compared to
the direct sum rule.  Thus, the predictions of the first derivative
sum rule are less reliable than those from the direct sum rule. At
zero density, the first derivative sum rule predicts a very large
$\rho$-meson mass and the continuum threshold is only about
$100\,\text{MeV}$ above the pole position.  Nevertheless, the first
derivative sum rule also predicts that the ratios $m^*_\rho/m_\rho$,
$F^*_\rho/F_\rho$, and $s_0^*/s_0$ all decrease as the density
increases, which agrees qualitatively with the predictions of the
direct sum rule.  The reasons for the failure of the first derivative
sum rule to reproduce the $\rho$-meson mass obtained from the direct
sum rule are discussed at length in Ref.~\cite{leinweber95}.

In the second derivative sum rule of Eq.~(\ref{sum-gen-n}) (with
$n=2$), the contributions of the quark and gluon condensates drop out
and the nonperturbative power correction starts with dimension six
condensates, the last term of the truncated OPE.  The perturbative
contribution and hence the continuum contribution is multiplied by an
extra factor of two relative to that for the direct sum
rule. Numerical analysis indicates that there is no Borel window where
the sum rule is valid. Thus, one cannot get information about
ground-state vector mesons from this sum rule.  This is also true for
the third and higher derivative sum rules, where there are no power
corrections at the level of the OPE truncation considered here.

All of the results above are for the $\rho$ and $\omega$ mesons.  A
similar analysis can be done for the $\phi$ meson. Again, we find the
same pattern. The direct sum rule gives the most reliable predictions,
the first derivative sum rule yields a less reliable result, and the
second and higher derivative sum rules are invalid. We find from the
direct sum rule $m^*_\phi/m_\phi=0.99\pm 0.01$ at nuclear matter
saturation density.  This rate of decrease is much smaller than that
for the $\rho$ and $\omega$ mesons. This is due to the dominance of
$m_s\langle\overline{s}s\rangle_{\rho_N}$ and its slow change with the
medium density.

\subsection{Discussion}

The Borel transform plays important roles in making the QCD sum-rule
approach viable. It suppresses excited state
contributions exponentially on the phenomenological side, thus
minimizing the continuum model dependence. It also improves OPE
convergence by suppressing
the high order power corrections factorially on the
QCD side. We observe that taking derivatives of the direct
sum rule with respect to $1/M^2$ is equivalent to a partial
reverse of the Borel transform.
It is thus not surprising to find that the continuum contribution
becomes more important and the convergence of the OPE becomes
slower in the derivative sum rules than in the direct sum rule.
Since the excited state contributions are modeled
roughly by a perturbative evaluation of the correlator
starting at an effective threshold, and the higher order OPE
terms are not well determined, there are more
uncertainties in the derivative sum rules
than in the direct sum rule.  In fact, a simultaneous fit of
both the direct and first derivative sum rules in vacuum reveals that
the first derivative sum rule plays a negligible role in
determining the fit parameters, when a $\chi^2$ measure weighted by
the OPE uncertainty and relative reliabilities of the sum rules
is used\cite{leinweber95}.

To improve the reliability of the derivative
sum rules, one must include more higher order terms in the OPE.
However, one usually does not have much
control of the values of the higher-dimensional condensates.
In addition, the derivative sum rule will always suffer from a
factorial enhancement of the terms contributing to the
continuum model. Therefore, the direct sum rule will always
yield the most reliable results for vector mesons.

Hatsuda and Lee invoked both the direct and the first derivative sum
rules\cite{hatsuda92}. Their results for the ratio $m_{\rho^*}/m_\rho$
are somewhat larger than those we obtained from the direct sum rule
but are somewhat smaller than those from the first derivative sum
rule.  This discrepancy is obviously attributed to their use of the
two sum rules simultaneously without weighing the relative merits of
the sum rules.  In Ref.~\cite{koike95}, Koike adopted both the first
and second derivative sum rules. His conclusion of slightly increasing
vector-meson masses in the medium depends on the use of the second
derivative sum rule, which we have found to be invalid when the OPE is
truncated at dimension six. Hatsuda {\it et al.}\cite{hatsuda95} also
pointed out some shortcomings of the second derivative sum
rule. However, their arguments are based on concerns over the lack of
information on the QCD side of the second derivative sum rule and the
absence of a ``plateau'' in the ratio of the second and first
derivative sum rules from which a mass is extracted.  In this paper,
we have extensively explored why these observations come about, and
why even Hatsuda {\it et al.}'s analysis is less than satisfactory.

The use of the ratio method is mainly driven by the expectation that
if the sum rules work well one should see a plateau in the predicted
quantities as functions of the Borel mass. The usual interpretation of
this criterion is that the ratio of {\it two} different sum rules,
proportional to certain spectral parameter of interest (e.g. mass),
should be flat as function of the Borel mass. Although it is true
ideally, this interpretation is potentially problematic in
practice. We have seen that the reliabilities and validities of two
sum rules are usually different. This feature cannot be revealed in
the ratio of the two sum rules. In addition, one can always achieve
the flatness of the ratio in the large Borel mass region, where both
sides of the sum rules are dominated by the continuum. However, one
learns little about the lowest pole in this case.

\begin{figure}[b]
\begin{center}
\epsfysize=11.6truecm
\leavevmode
\setbox\rotbox=\vbox{\epsfbox{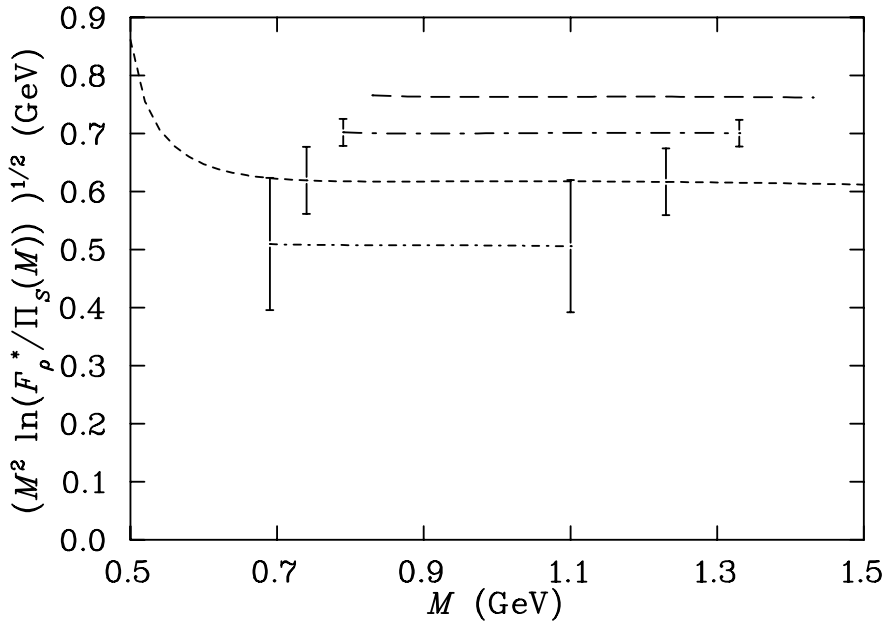}}\rotl\rotbox
\end{center}
\caption{The $\rho$-meson mass as obtained from the right-hand side of
Eq.~(\protect{\ref{mv-borel}}) as a function of $M$ at various nuclear
matter densities. The curves from top down correspond to $\rho_N=0$,
$\rho_N=0.5\,\rho_N^0$, $\rho_N=\rho_N^0$, and
$\rho_N=1.5\,\rho_N^0$. The error bars denoting the valid regimes are
obtained from the relative errors of Fig.~\protect{\ref{fig-1}}. Note
how the Borel regime shifts and becomes smaller as the density
increases.  The curve for $\rho_N=\rho_N^0$ is plotted outside the
valid region to demonstrate the importance of carefully selecting a
Borel region. }
\label{fig-6}
\end{figure}

To make contact with the plateau criterion, we propose that if a sum
rule works well, one should see a plateau in the plot of an extracted
quantity expressed as a function of the Borel mass. For example, from
the direct sum rule of~(\ref{sum-gen-0}), one can express the
vector-meson mass as
\begin{equation}
m^*_{\rm v}=\left[ M^2\, \ln\, \left({F^*_{\rm v}\over \Pi_S(M)}
\right)\right]^{1 / 2}\ ,
\label{mv-borel}
\end{equation}
where $\Pi_S(M)$ denotes the right-hand side of Eq.~(\ref{sum-gen-0}).
In Fig.~\ref{fig-6}, we plot the right-hand side of
Eq.~(\ref{mv-borel}) for the $\rho$-meson case at four different
densities, with optimized values for $F_{\rm v}^*$ and $s^*_0$. One
indeed sees very flat curves within the region of validity denoted by
the error bars.  It should be emphasized that (1) this interpretation
only involves one sum rule and is thus different from the ratio method
previously used in the literature; (2) the value of $m_{\rm v}^*$ is
only meaningful in the valid region of a sum rule; (3) in this
interpretation, the plateau criterion is a true criterion, measuring
the quality of the overlap between the two sides of a valid sum
rule. For curiosity, we also display the curve outside the validity
region for the case of $\rho_N=\rho^0_N$ in Fig.~\ref{fig-6}. One
notices a deviation from the plateau just outside the valid Borel
region. This feature supports our selection of $10\%$ as the criterion
for OPE convergence.

In the present analysis as well as the previous works, the linear
density approximation has been assumed for the in-medium condensates.
For a general operator there is no systematic way to study
contributions that are of higher order in the medium density.
Model-dependent estimates in Ref.~\cite{cohen92} suggest that the
linear approximation to $\langle\overline{q}q\rangle_{\rho_N}$ should
be good (higher-order corrections $\sim 20\%$ of the linear term) up
to nuclear matter saturation density. In our analyses, we have
assigned generous uncertainties to various condensate values and
parameters. We expect these will cover the uncertainties arising from
the linear density approximation. As the medium density increases, the
deviation from the linear density approximation will increase. One
then needs more precise knowledge of the density dependence of various
condensates in order to have reliable QCD sum-rule predictions.

   As in previous works, we have neglected the dimension-six
twist-four operators in (\ref{rho-coef}), as the nucleon matrix
elements of these operators are unknown.  However, estimates may be
obtained from deep-inelastic-scattering data provided one is willing
to make a few additional assumptions \cite{hatsuda95}.  Taking the
estimates given in Ref.\ \cite{hatsuda95}, we find these additional
contributions have little effect on our results.  At saturation
density, our present in-medium $\rho$-meson mass of 0.59 GeV is
increased slightly to 0.62 GeV.  Taking a 100\% uncertainty on the
twist-four contributions has no apparent effect on the present
uncertainty of 0.11 GeV.  At saturation density, the ratio
$m_{\rho^*}/m_\rho$ is shifted from 0.78 to 0.82 which is small
relative to the uncertainty of 0.08.  Certainly further study of the
twist-four contributions is required.  However, we do not expect such
contributions to significantly alter the conclusions of this paper.

As a final remark, we comment on the electromagnetic width
of the $\rho$ meson, $\Gamma (\rho^0\rightarrow e^+ e^-)$.
In free space, it is given by\cite{bhaduri88,svz79}
\begin{equation}
\Gamma (\rho^0\rightarrow e^+ e^-)={1\over 3}\,\alpha^2_E\,
m_\rho\,\left({4\pi\over g_\rho^2}\right)
={4\pi\over 3}\,\alpha^2_E\, \left({F_\rho\over m_\rho}\right)\ ,
\end{equation}
where $\alpha_E$ is the electromagnetic coupling constant.
The modification of this result in medium may be estimated by
replacing $m_\rho$ and $F_\rho$ with their corresponding
values in medium. The ratio of the free space and in-medium
widths can be expressed as
\begin{equation}
{\Gamma^* (\rho^0\rightarrow e^+ e^-)\over
\Gamma (\rho^0\rightarrow e^+ e^-)}=
{m_\rho\over m^*_\rho}\, {F^*_\rho\over F_\rho}\ .
\end{equation}
Note that $m_\rho/ m^*_\rho$ increases while
$F^*_\rho/ F_\rho$ decreases with increasing
density. However, the rate of decrease for $F^*_\rho/ F_\rho$
is larger than the rate of increase for  $m_\rho/ m^*_\rho$.
Consequently, the ratio for the widths is
less than $1$. At $\rho_N=\rho^0_N$, our estimate is
$\Gamma^*/\Gamma=0.85\pm 0.10$.
This implies that the $\rho$-meson
electromagnetic width becomes smaller in nuclear matter.
This behavior might be observed in the proposed experiment
 studying dileptons as a probe of vector mesons in the dense
and hot matter\cite{dil}.

\section{Conclusion}
\label{conclusion}

In this paper, we have carefully examined the QCD sum rules for vector
mesons in nuclear matter. Our primary concern has been on the validity
and reliability of various sum rules. We emphasize that the sum rules
do not work equally well due to the truncation of the OPE and the use
of a model for the phenomenological spectral density. In particular,
the derivative sum rules are less reliable than the direct sum
rule. This is attributed to: (1) the perturbative contribution and
hence the continuum contribution become increasingly important in the
derivative sum rules; (2) the high order terms in the OPE become
increasingly important in the derivative sum rules; (3) part (or all)
of the nonperturbative information is lost in the derivative sum
rules.  We therefore conclude that any predictions based on (or
partially on) second or higher derivative sum rules are incorrect
given the level of the OPE truncation adopted in the literature. One
should avoid using the derivative sum rules in practical applications.

We tested this conclusion numerically by analyzing the sum rules with
regard to pole dominance and OPE convergence\cite{leinweber90}.  A
Monte Carlo based error analysis was used to provide reliable
uncertainties on our predictions and remove the sensitivity of the
results to the input parameters\cite{leinweber95}.  We found that the
direct sum rule satisfies our criteria and leads to reliable
predictions. The first derivative sum rule suffers from a small region
of validity and large continuum contributions throughout. The second
and higher derivative sum rules are invalid.

Our analysis confirms that the QCD sum-rule approach  predicts
a decrease of vector-meson masses with increasing density,
and resolves the debate between Hatsuda {\it et al.}\cite{%
hatsuda95} and Koike\cite{koike95}.
The prediction of a slight increase of vector-meson masses in medium
is based on an invalid second derivative sum rule.

We note that all previous authors have used the ratio method in the
analysis of the sum rules, which has many drawbacks and may lead to
incorrect results. We encourage the community to adopt the approach
developed in Ref.~\cite{leinweber95} which checks the quality of the
overlap between two sides of each individual sum rule by monitoring
pole dominance and the convergence of the OPE. This approach also
allows one to realistically estimate the uncertainties and reveal the
predictive ability of QCD sum rules.

The analysis presented here is the most reliable QCD analysis
of in-medium vector-meson properties. At nuclear matter saturation
density, we predict
\begin{equation}
{m_\rho^*\over m_\rho}=0.78\,\pm\,0.08,\hspace{1cm}
{\Gamma^* (\rho^0\rightarrow e^+ e^-)\over
\Gamma (\rho^0\rightarrow e^+ e^-)}=0.85\,\pm\,0.10\ ,
\end{equation}
and look forward to experimental vindication of these
results\cite{dil}.

\acknowledgements

We would like to thank Tom Cohen for insightful conversations. This
work was supported by the Natural Sciences and Engineering Research
Council of Canada.



\begin{references}
\bibitem{dil} The Phoenix collaboration at RHIC and the Hades
       collaboration at GSI will look for this effect.
\bibitem{mentioned}This point was also noted in Ref.\
\protect\cite{hatsuda95}.
\end{references}
\end{document}